\documentclass[journal]{IEEEtran}

\usepackage{multirow}

\ifCLASSINFOpdf
  \usepackage{graphicx}
 \usepackage{epstopdf}
  \graphicspath{{./pdf/}{./jpeg/}{./eps/}}
  \DeclareGraphicsExtensions{.pdf,.jpeg,.png,.eps}

\ifCLASSOPTIONcompsoc
  \usepackage[caption=false,font=normalsize,labelfont=sf,textfont=sf]{subfig}
\else
  \usepackage[caption=false,font=footnotesize]{subfig}
\fi

\usepackage{bytefield}
\usepackage{color}
\usepackage{url}
\usepackage{balance}
\begin{document}

\title{KCN: Knowledge Centric Networking}

\author{
	\IEEEauthorblockN{
		Marinos Charalambides\IEEEauthorrefmark{1},
		Daphne Tuncer\IEEEauthorrefmark{2},
		Ning Wang\IEEEauthorrefmark{3},
		George Pavlou\IEEEauthorrefmark{1}\\
	}
	\IEEEauthorblockA{
		\IEEEauthorrefmark{1}University College London, UK
	}	
	\IEEEauthorblockA{
		\IEEEauthorrefmark{2}Imperial College London, UK
	}	
	\IEEEauthorblockA{
		\IEEEauthorrefmark{3}University of Surrey, UK
	}
}

\maketitle

\begin{abstract}
The advent of multi-domain and multi-requirement digital services requires an underlying network ecosystem able to understand service-specific contexts. In this work, we propose Knowledge Centric Networking (KCN), a paradigm in which knowledge is positioned at the center of the networking landscape. KCN enables approaches by which in-network knowledge generation and distribution can be used to support advanced network control intelligence that is essential to handle complexity and uncertainty in these emerging digital services. In this paper, we introduce the principles of KCN and present an architecture for its realization that enables in-network knowledge creation, distribution, storage and processing, both within a single domain/player and across potentially heterogeneous domains/players. 
\end{abstract}

\begin{IEEEkeywords}
Networking paradigm; network intelligence; knowledge management.
\end{IEEEkeywords}

\IEEEpeerreviewmaketitle

\section{Introduction}
\label{sect:Intro}

The digitation of all major sectors is imposing stringent requirements on the networking infrastructures that need to be in place in order to transport an increasing volume of content, connect a wide range of heterogeneous devices and support a growing number of ``smart" and adaptive services. To handle the increasing complexity associated with deploying, maintaining and managing these infrastructures, knowledge needs to be positioned at the center of the networking landscape.

While the concept of a \textit{knowledge plane} was proposed more than a decade ago \cite{clarkSigcomm03} and knowledge was a key feature of autonomic networking architectures \cite{kephartComp03}\cite{schmidAuto06}, such frameworks have never been realized. Furthermore, current practices in network resource management are usually based on centralized offline systems, which optimize network performance over long timescales. As such, the monitoring mechanisms in place operate at infrequent intervals generating global system views. For these reasons, configuration decisions are long lived given the absence of knowledge on instantaneous network conditions, which lead to suboptimal use of resources and user experience. In addition, the knowledge acquired by a stakeholder tends to stay within the boundaries of that stakeholder due to legal issues and market competition. This however prevents stakeholders of taking advantage of potentially useful knowledge that could facilitate improved configuration decisions and hence provide stricter expected levels of operation guarantees under changing conditions. 

The emergence of novel networking paradigms such as Mobile Edge Computing and Information Centric Networking in the last ten years have stressed the need for sophisticated in-network intelligence in order to undertake complex tasks.  A key requirement is that of adaptive resource control as a response to changing demand and network conditions. It has thus become obvious that the current practice of a relatively simple network infrastructure with an ever-increasing complexity of over the top “third party” management systems will be no longer sustainable in the context of emerging and future networked applications.

In this article we propose Knowledge-Centric Networking (KCN), an approach in which both service-oriented and network-oriented knowledge could be harvested, stored, processed and also shared by different stakeholders. In KCN, knowledge is viewed as processed information that creates awareness of the surrounding environment and which ultimately enables the network to react to emerging conditions in an automated and timely manner. Two degrees of knowledge can be identified: (i) facts derived from basic information processing, e.g. information aggregation and filtering, that represent snapshots of the system state; and (ii) advanced knowledge derived from algorithmic processing and reasoning over facts, e.g. the outcome of prediction algorithms, inference logic and learning. Building such knowledge about the network itself, the services it supports and their users will enable self-adaptation and help achieve close-to-optimal performance under dynamic, continuously changing conditions. This will subsequently enable more demanding applications that are not possible today given the relatively unpredictable nature of the current Internet.

\section{The KCN Architecture}
\label{sect:KCNarch}

Over the recent years the concept of Software-Defined Networking (SDN) has been materialized as an architectural model for the operation and management of today's networks, with deployment covering data centers to telecom operator \cite{kimCommag13}\cite{laraSurvey13} and even Internet Exchange Point (IXP) \cite{bruyereJSAC18} infrastructures. The acceleration of SDN deployment as a new model of network architecture has further been reinforced by the evolution of network services through Network Functions Virtualization (NFV) \cite{mijumbiCommag16} that support decoupling of the network control and data planes. SDN offers an ideal solution to realize the KCN paradigm given that knowledge-driven, programmable control functions can be flexibly organized and enforced wherever needed, without the need to be physically coupled with the data plane functions as is the case today. However, instead of following the standard master-slave style of SDN operation where all the knowledge and control intelligence is realized at a central controller site, we advocate a novel distributed knowledge ecosystem in which necessary knowledge is distributed at the right locations for easing network control. 

\begin{figure*}[t]
\centering{
\includegraphics[trim = 0cm 0cm 0cm 0cm, clip,scale=0.45]{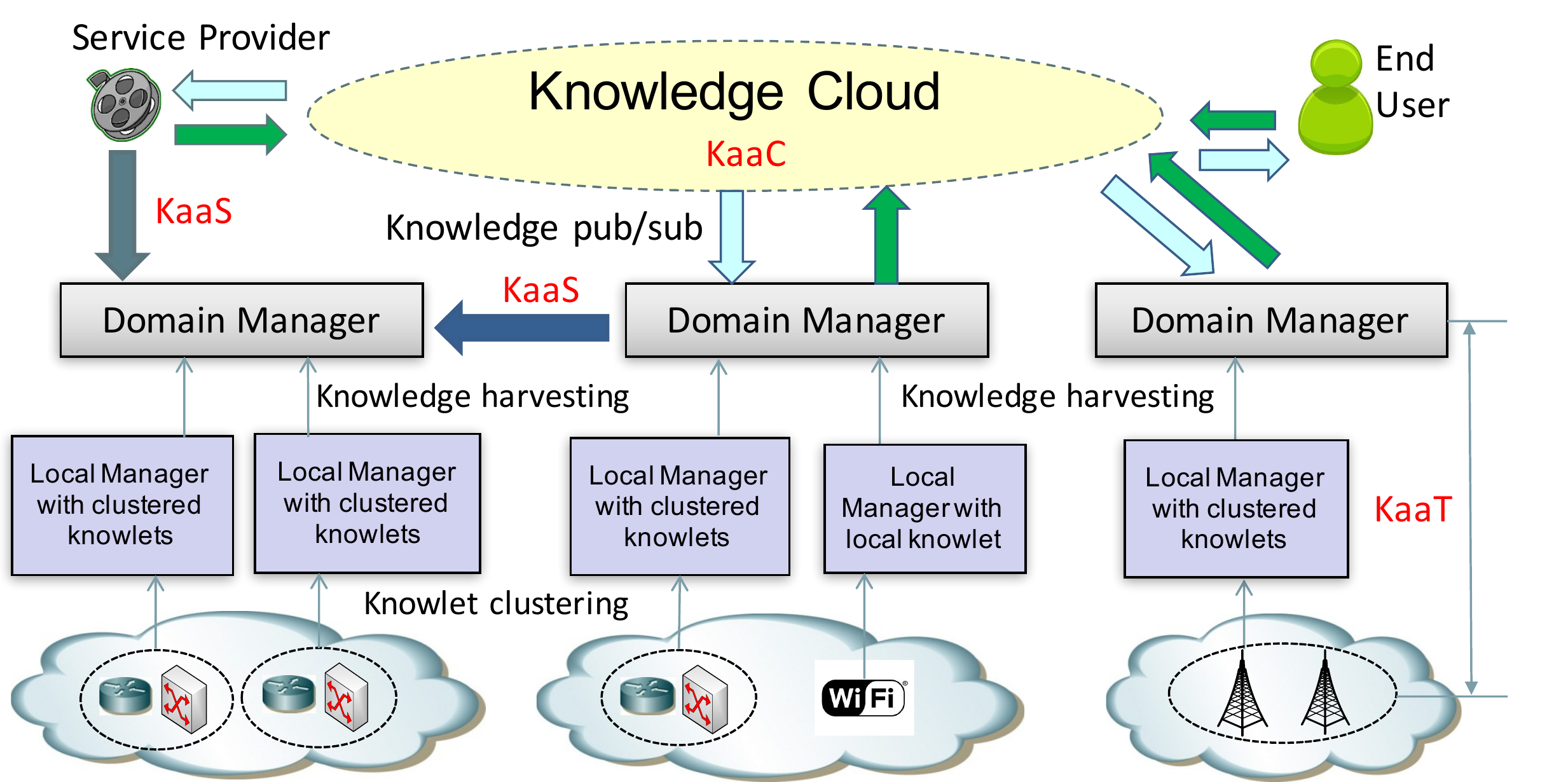}  
}
\caption{Knowledge Centric Networking Architecture.}
\label{fig:kcnArch}
\vspace{-5mm}
\end{figure*}

Fig. \ref{fig:kcnArch} presents the proposed SDN-based KCN architecture that realizes a knowledge ecosystem with three levels of hierarchical knowledge structure. First of all, pieces of local knowledge at the device level, called \textit{knowlets}, are individually collected, or are locally clustered based on a subset of network elements with higher degree of operation interdependence. A local SDN manager for a device or a cluster is responsible for collecting/processing local knowlets and reporting to the central domain manager. Such a local manager can also take and enforce local decisions for controlling the device/cluster behavior based on local knowledge, without a global network view but within centrally predefined limits in order to guarantee convergence and stability. The second level refers to knowlet harvesting by the central domain manager from all the local SDN managers in the domain. The objective is for the central manager to build domain-level knowledge which can be used for controlling the network behavior as a whole. This includes both facts and advanced knowledge through necessary prediction, inference and learning.

In addition to capturing and exploiting knowledge within a single administrative domain, the proposed architecture supports knowledge sharing between different stakeholders in the Internet marketplace for achieving coordinated end-to-end resource management. We investigate the following three styles of knowledge exchange in the proposed ecosystem:

\begin{itemize}
	\item \textit{\textbf{Knowledge as a Tool (KaaT)}}, in which knowledge is used to support intelligent and adaptive service/network control within a single administrative domain, \textit{e.g.,} an ISP or cellular network;
	\item \textit{\textbf{Knowledge as a Service (KaaS)}}, in which knowledge could be explicitly transferred from a knowledge provider to a knowledge consumer as a commercial service based on Service Level Agreements (SLAs), in order to generate new business opportunities and enable coordinated resource control across different domains;
	\item \textit{\textbf{Knowledge as a Cloud (KaaC)}}, in which a group of stakeholders with mutual incentives may flexibly share their knowledge through a common knowledge cloud in an open peer-to-peer fashion, without SLAs.
\end{itemize}

Knowledge clustering and harvesting by the domain manager within a domain adheres to the KaaT style of knowledge exchange \textit{i.e.,} knowledge used as a tool for enabling better network control. However, knowledge may also be exchanged horizontally, between adjacent domains / homogeneous stakeholders, or vertically between heterogeneous stakeholders, \textit{e.g.,} between a service and network provider; this is the KaaS client-server style, with knowledge offered as a service to another stakeholder. Finally, knowledge may be also placed in a knowledge cloud and accessed by different stakeholders with common interests, which is the KaaC peer-to-peer style. In both KaaS and KaaC, a domain performs network control based not only on its own knowledge but also on knowledge related to other domains. All these styles of interaction are depicted in Fig. \ref{fig:kcnArch}. We elaborate further on them in the next section.

\section{Knowledge as a Tool/Service/Cloud}
\label{sect:Kaa*}

\subsection{Knowledge as a Tool}
\label{sect:KaaT}

Emerging network infrastructures are increasingly service and content-aware, and as a result the underlying network is able to capture context information on the service operated and content being transmitted. This is in contrast to the current content/service agnostic networks for which only limited context information can be obtained, usually concerning only the network itself, \textit{e.g.,} link/node conditions. KaaT involves deriving and harvesting/disseminating knowledge within a single domain, supported by the presented hierarchical SDN infrastructure with the involvement of local managers at device/cluster level and the central domain manager. The fundamental requirement is for the local managers to build necessary knowlets based on raw context information, which are sufficient for them to perform locally designated control functions at the device/cluster level. This approach can reduce substantially the knowledge processing overhead at the domain manager side while prediction, machine learning and knowledge inference techniques are deployed at different locations in the SDN hierarchy. Examples include prediction techniques that can be employed to determine content consumption patterns and infer content popularity trends \cite{szaboComms10}, and also learning mechanisms that can classify observed user behavior \cite{lerouxAmbient13} and dynamically adapt configuration decisions accordingly \cite{corapiAI09}\cite{tangariTMA20}.

\subsection{Knowledge as a Service}
\label{sect:KaaS}

KaaS, in a similar fashion to KaaC, involves multiple stakeholders in the KCN ecosystem. A knowledge provider may disseminate its own knowledge in order to provide relevant opportunities to interested knowledge consumers. For example, aggregated knowledge on user behavior derived from the profiles of networked content consumption can be of common interest to both network operators and content providers \textit{e.g.,} \cite{claeysJNCA16}. However, no single stakeholder is able to capture this knowledge without relying on information provided by others. The sharing of such knowledge between the stakeholders can offer mutual benefits: for instance, with comprehensive knowledge on content popularity/availability and user location distributions with associated mobility patterns, network operators may strategically cache content at routers, gateways or even cellular base-stations in order to efficiently reduce content traffic within the network \cite{geMultimedia17}. In return, this will also benefit content providers for optimizing their server loads. In addition, knowledge sharing between network infrastructure providers will enable coordinated resource sharing/trading with each other. 

The approach for supporting KaaS (and KaaC) is similarly based on the proposed SDN-based architecture, with the central manager at each domain being also responsible for knowledge exchange with other domains, in addition to local decision making and enforcement. Apart from the communication mechanisms for enabling such knowledge transfer,  behavioral interactions when transferring ``dynamic" knowledge are very important. For example, the frequency of knowledge updates can play a vital role on the overall system performance in terms of efficiency and stability. An interesting KaaS scenario is bi-directional knowledge sharing between two stakeholders. In this case, the decision made by one of them should be based on both the local knowledge and the most up-to-date knowledge disseminated by the other stakeholder. The outcome of the local decision may generate new knowledge, which may in turn impact decision-making on the other side. As such, ``synchronization" of knowledge sharing is an important research challenge, addressing what and how frequently to transfer by the provider in order to achieve stability. This interaction can be modeled as a ``game" in the knowledge ecosystem, with different players jointly optimizing their objectives based on cooperative game theory, \textit{e.g.,} \cite{jiangSig09}.

\subsection{Knowledge as a Cloud}
\label{sect:KaaC}

In comparison to KaaS, KaaC exhibits very different characteristics with respect to the knowledge dissemination and consumption style. Here, there are no knowledge SLAs but knowledge is instead exchanged in an open peer-to-peer fashion: knowledge providers do not care about the exact network location and identity of the consumers accessing their published knowledge, while the latter are only interested in the knowledge itself rather than the network location where it comes from. Given this style of interaction, a publish/subscribe approach can be used for realizing this ``knowledge cloud". In such an open knowledge ecosystem, any player may act either as a provider/publisher, a consumer/subscriber or both. On the other hand, knowledge access control mechanisms may be put in place. It can be achieved through policy-based control (\textit{e.g.,} scoping, filtering) over the knowledge published in the cloud.

The publish/subscribe nature of the knowledge cloud makes an approach based on the concept of ICN \cite{xylomenosSurvey14} an ideal candidate for realizing such open knowledge sharing infrastructure. Given the fact that a lot of the relevant information can be produced by mobile end-user devices and also considering aspects such as design simplicity, scalability and ease of deployment, an ICN-oriented Custodian-Based Information Sharing (CBIS) approach \cite{jacobCommag12} can be used for building the KaaC knowledge sharing platform. This will allow a large number of end users taking the roles of knowledge providers and consumers to flexibly share knowledge whenever there is network connectivity, but without relying on a dedicated physical infrastructure. This is particularly beneficial for end users on-the-move to publish their knowledge in an opportunistic fashion with any available network connection, \textit{e.g.,} their quality of experience (QoE) that can be used by a network operator to make necessary adaptations. The CBIS-based approach will facilitate this type of operation through indirection and late-binding \textit{i.e.,} a series of entities (knowledge custodians) can be dynamically defined, responsible for collecting, replicating and disseminating knowledge, utilizing the currently available communication facilities. In addition, the prefix-based name space proposed by CBIS can support high scalability for necessary aggregation of the published knowledge when shared \textit{e.g.,} dynamicity patterns of content popularity or user density at different levels of geographical coverage. 

SDN can facilitate the management of knowledge custodians by first enabling the aggregation and processing of knowledge at domain managers, which can subsequently act as domain-level knowledge custodians interfacing with their counterparts in different domains. In addition, it is also possible that central managers delegate knowledge aggregation and processing tasks to their local SDN managers, which can autonomously exchange knowlets across domains for local optimization, but without always getting the respective domain managers involved.

\section{Conclusions}
\label{sect:conclusions}

This paper presents \textit{Knowledge Centric Networking}, a novel paradigm whose key objective is to enable in-network knowledge generation and distribution in support of network control and management. A KCN-based ecosystem provides the required features to equip the next generation Internet with the necessary intelligence for handling complex requirements under dynamic conditions. Such an ecosystem, seamlessly coupled with the SDN architecture, can gracefully support the ever increasing complexity and heterogeneity of future networked services. The paper describes three styles of knowledge exchange based on SDN principles: Knowledge as a Tool (KaaT), Knowledge as a Service (KaaS) and Knowledge as a Cloud (KaaC); and discusses example scenarios illustrating the role of these different types of knowledge.

\section*{Acknowledgment}
This research was funded by the EPSRC KCN project (EP/L026120/1).

\end{document}